\documentclass[11pt]{amsart} 
\usepackage{times,amssymb,amsmath,float,nicefrac} 
\usepackage{euscript,graphics}
\usepackage[dvips]{epsfig}

\evensidemargin \oddsidemargin
\newtheorem{theorem}{Theorem}

\newtheorem{lemma}[theorem]{Lemma}

\renewenvironment{proof}[1]{\noindent {\it Proof~:} #1}
{\ \rule{1mm}{2mm}\medskip}

\renewcommand\qed{\rule{1mm}{2mm}\medskip}

\newcommand{\remove}[1]{}

\newcommand\nd{\noindent}

\def\Vol{\qopname\relax{no}{Vol}}

\newcommand\nc\newcommand
\nc\bfa{{\mathbf a}}\nc\bfA{{\mathbf A}}\nc\cA{{\mathcal A}}
\nc\bfb{{\mathbf b}}\nc\bfB{{\mathbf B}}\nc\cB{{\mathcal B}}
\nc\bfc{{\mathbf c}}\nc\bfC{{\mathbf C}}\nc\cC{{\mathcal C}}
\nc\bfd{{\mathbf d}}\nc\bfD{{\mathbf D}}\nc\cD{{\mathcal D}}
\nc\bfe{{\mathbf e}}\nc\bfE{{\mathbf E}}\nc\cE{{\mathcal E}}
\nc\bff{{\mathbf f}}\nc\bfF{{\mathbf F}}\nc\cF{{\mathcal F}}
\nc\bfg{{\mathbf g}}\nc\bfG{{\mathbf G}}\nc\cG{{\mathcal G}}
\nc\bfh{{\mathbf h}}\nc\bfH{{\mathbf H}}\nc\cH{{\mathcal H}}
\nc\bfi{{\mathbf i}}\nc\bfI{{\mathbf I}}\nc\cI{{\mathcal I}}
\nc\bfj{{\mathbf j}}\nc\bfJ{{\mathbf J}}\nc\cJ{{\mathcal J}}
\nc\bfk{{\mathbf k}}\nc\bfK{{\mathbf K}}\nc\cK{{\mathcal K}}
\nc\bfl{{\mathbf l}}\nc\bfL{{\mathbf L}}\nc\cL{{\mathcal L}}
\nc\bfm{{\mathbf m}}\nc\bfM{{\mathbf M}}\nc\cM{{\mathcal M}}
\nc\bfn{{\mathbf n}}\nc\bfN{{\mathbf N}}\nc\cN{{\mathcal N}}
\nc\bfo{{\mathbf o}}\nc\bfO{{\mathbf O}}\nc\cO{{\mathcal O}}
\nc\bfp{{\mathbf p}}\nc\bfP{{\mathbf P}}\nc\cP{{\mathcal P}}
\nc\bfq{{\mathbf q}}\nc\bfQ{{\mathbf Q}}\nc\cQ{{\mathcal Q}}
\nc\bfr{{\mathbf r}}\nc\bfR{{\mathbf R}}\nc\cR{{\mathcal R}}
\nc\bfs{{\mathbf s}}\nc\bfS{{\mathbf S}}\nc\cS{{\mathcal S}}
\nc\bft{{\mathbf t}}\nc\bfT{{\mathbf T}}\nc\cT{{\mathcal T}}
\nc\bfu{{\mathbf u}}\nc\bfU{{\mathbf U}}\nc\cU{{\mathcal U}}
\nc\bfv{{\mathbf v}}\nc\bfV{{\mathbf V}}\nc\cV{{\mathcal V}}
\nc\bfw{{\mathbf w}}\nc\bfW{{\mathbf W}}\nc\cW{{\mathcal W}}
\nc\bfx{{\mathbf x}}\nc\bfX{{\mathbf X}}\nc\cX{{\mathcal X}}
\nc\bfy{{\mathbf y}}\nc\bfY{{\mathbf Y}}\nc\cY{{\mathcal Y}}
\nc\bfz{{\mathbf z}}\nc\bfZ{{\mathbf Z}}\nc\cZ{{\mathcal Z}}
\nc\od{{\bar d}}\nc\ow{{\bar w}}\nc\odelta{{\bar\delta}}
\nc\ox{{\bar x}}\nc\oy{{\bar y}}\nc\ou{{\bar u}}
\nc\oh{{\bar h}}

\newcommand\reals{{\mathbb R}}

\nc\dgv{\delta_{\text{\rm GV}}}
\nc\rcrit{R_{\text{\rm crit}}}
\nc\Esp{E_{\text{\rm sp}}}
\renewcommand\epsilon{\varepsilon}

\newcommand{\beeq}{\begin{eqnarray*}}
\newcommand{\eneq}{\end{eqnarray*}}

\newcommand{\half}{\nicefrac12}

\begin{document}
\title[A bound on Grassmannian codes]
{A bound on Grassmannian codes}

\author[A. Barg]{A. Barg$^\ast$}\thanks{$^\ast$
Dept. of ECE, University of Maryland,
  College Park, MD 20742.
Supported in part by NSF grants CCR0310961,
CCF0515124.}
\author[D. Nogin]{D. Nogin$^\dag$}\thanks{$^\dag$ 
IPPI RAN, Bol'shoj Karetnyj 19, Moscow 101447, Russia.
Supported in part by NSF grant CCR0310961 and RFFI grant 02-01-22005.}

\begin{abstract} We give a new asymptotic
upper bound on the size of a code
in the Grassmannian space. The bound is better than the upper bounds
known previously in the entire range of distances except
very large values.
\end{abstract}
\maketitle

\subsection*{1. Introduction.} Let $G_{k,n}(\reals)$ be the Grassmann
manifold, i.e., the set of $k$-planes passing through the origin in
$\reals^n$. 
Our focus is the packing problem in $G_{k,n}$, i.e., the problem of
estimating the number of planes whose pairwise distances are
bounded below by some given value $\delta$, for a suitably defined distance
function $d(p,q).$ This problem has attracted attention in the
recent years for several reasons. As a coding-theoretic (geometric)
problem, it is a natural generalization of the coding problem for
the projective space $P\reals^{n-1}$ and a closely related case of the
sphere in $\reals^n$, both having long history in coding theory
\cite{con96}.
This problem arises also in engineering applications related to transmission
of signals with multiple antennas in wireless environment \cite{agr01}.
Finally, \cite{sho98} introduced a construction of Grassmannian packings 
which is closely related 
to the construction of quantum stabilizer codes, another
subject of interest in recent years.

There are several possibilities to define a metric on $G_{k,n}$ \cite{ede99}.
We consider the so-called chordal metric (projection 2-norm
in the terminology of \cite{ede99}), which can be defined in two
equivalent ways. 
By a well-known fact \cite{jam54}, given two planes $p, q\in G_{k,n}$ 
one can define $k$ principal angles between them.
This is done recursively as follows: take unit vectors $x_1\in p,y_1\in q$
with the maximum possible angular separation and denote this angle by 
$\theta_1.$ In step $i=2,\dots,k,$ take the unit vectors
$x\in p, x_i\perp\langle x_1,\dots,x_{i-1}\rangle$ and $y\in q,
y_i\perp\langle
y_1,\dots,y_{i-1}\rangle$ with the maximum possible angle between them
and denote this angle by $\theta_i$. In this way we
obtain the set of principal angles $0\le \theta_k\le\dots\le\theta_1
\le \nicefrac\pi2$; moreover, $(x_1,\dots,x_k)$ and $(y_1,\dots,y_k)$ form
orthonormal bases in $p$ and $q$, respectively.

Let $\sin\theta=(\sin\theta_1,\dots,\sin\theta_k)$.
For a matrix (vector) $P$ let $\|P\|=\sqrt{\sum_{i,j}P_{ij}^2}$ denote
its Euclidean 2-norm. 
Define the {\em chordal distance} between
$p$ and $q$ as follows: $d(p,q)=\|\sin\theta\|.$ 
It turns out \cite{con96} that the
Grassmannian space with the chordal metric affords an isometric embedding
in a sphere $S_r$ of radius $r=\sqrt{k(n-k)/n}$ in $\reals^{(n-1)(n+2)/2}$.
To describe it, let $A_p$ be a ``generator matrix'' of $p$,
i.e., a $k\times n$  matrix whose rows form an orthonormal basis of $p$. 
Then the orthogonal projection from $\reals^n$
on $p$ can be written as $\Pi_p=A_p^t A_p$.
Define a map $\Phi: G_{k,n}\to S_r$ as $\Phi(p)=\Pi_p-\nicefrac kn I_n$
(the plane is mapped to the traceless part of the projection on it).
For any $p$, the norm of $\Phi(p)$ equals 
$\|\Pi_p-\nicefrac kn I_n\|=r.$  The main
result of \cite{con96} is that the mapping $\Phi$ is an isometry
in the sense that 
  \begin{equation}\label{eq:isom}
    d^2(p,q)=\half\|\Pi_p-\Pi_q\|^2.
  \end{equation}

We call a collection of $M$ points in $G_{k,n}$ with pairwise
distances at least $\delta$ an $(M,\delta)$ {\em code} 
in the Grassmannian
space and call $\delta$ the distance of the code.
By (\ref{eq:isom}) such a code gives rise to an $(M,\sqrt 2\delta)$ code 
$\cC\subset S_r$, so any upper bound on the distance of $\cC$ gives an
estimate on the distance of $G_{k,n}$. In particular, by Rankin's bounds
\cite{ran55} for any $(M,\delta)$ code $\cG,$
  $$
\delta\le \begin{cases} \frac{k(n-k)}n
\frac{M}{M-1} &\mbox{if }
M\le n(n+1)/2,\\ \frac{k(n-k)}n&\mbox{if } M>
n(n+1)/2.
\end{cases}
  $$
These bound are tight in the sense that there
exist codes that meet them with equality \cite{sho98,con96}.
However, in the majority of cases, particularly, for codes of the
large size,
direct application of bounds on spherical
codes to codes in $G_{k,n}$ gives poor results because the image of $G_{k,n}$ 
on $S_r$ forms a very sparse subset of it.

\begin{figure}
\centering
\includegraphics[width=4.8cm]{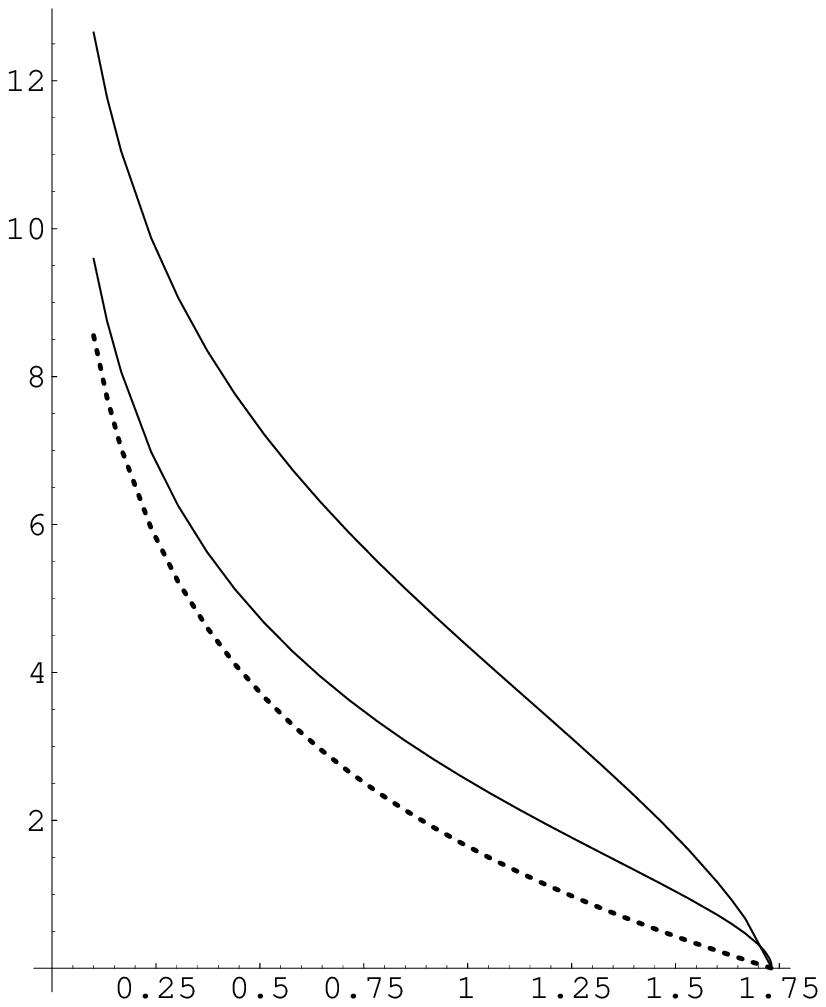}
\put(0,0){\text{\small $\delta$}}
\put(-150,160){\text{\small $R$}}
\caption{Bounds on $R(\delta)$ for $k=3$ (top to bottom): upper estimates
$R_{\text{LP}}$ \cite{bac05}, $R_{\text R}$ (Theorem 1); 
lower bound $R_{\text{GV}}$ \cite{bar02a}.}
\end{figure} 

We will be concerned with asymptotic bounds on $M$ for a given value
of the code distance $\delta$. 
Let $R=R(\delta)=\limsup_{n\to\infty} (1/n)\ln M$
be the largest possible {\em rate} of a sequence of codes 
with distance $\delta$ in $G_{k,n}$.
It is possible to compute the volume bounds on $R$ analogous
to the Gilbert-Varshamov and Hamming bounds of coding theory \cite{Lin99a}. 
Namely, it is proved in \cite{bar02a}
that for all $0\le\delta\le \sqrt k,$
  \begin{equation}\label{eq:gv-hamm}
 R_{\text{GV}}(\delta)\le R\le R_{\text H}(\delta),
  \end{equation}
where
    \begin{align*}
   R_{\text{GV}}(\delta)&=- k\ln (\delta/\sqrt k), \\
   R_{\text H}(\delta)&=- k \ln\bigg(\sqrt{1-\sqrt{1-
            \frac{\delta^2}{2k}}}\bigg).
  \end{align*}
The upper bound was subsequently improved in \cite{bac05} relying
on Delsarte's linear programming method in the form developed in
\cite{kab78}. The result of \cite{bac05} is
as follows:
  $$
   R\le R_{\text{LP}}(\delta):=k[(1+s)\ln(1+s)-s\ln s], 
  $$
where $s=(\nicefrac k2)\big(\frac{\sqrt k}\delta-1\big).$ 
This bound 
coincides with the result of \cite{kab78} for $k=1$ (codes in the 
projective space) and can be viewed as its generalization.
For $k=1$, $R_{\text{LP}}(\delta)< R_{\text H}(\delta)$
for all $0<\delta\le 1.$ However, for greater $k$ the inequality
$R_{\text{LP}}(\delta)< R_{\text H}(\delta)$ holds only
for $\delta$ close to $\sqrt k$ and thus the linear programming
bound provides a better estimate of $R$ only for large values
of the distance. For instance, for $k=2,3$ the crossing point is
$\delta=0.74,1.31$ respectively. We note that $R_{\text{LP}}(\sqrt k)=0$
showing that the lower bound $R_{\text{GV}}$ is tight for $\delta=\sqrt k$.
In this note we establish an improved upper bound stated in the following
theorem.
\begin{theorem} 
           $$R\le R_{\rm R}(\delta):=- k \ln\bigg(\sqrt{1-\sqrt{1-
            \frac{\delta^2}{k}}}\bigg).$$
\end{theorem}
Clearly $R_{\rm R}(\delta)<R_{\text H}(\delta)$ for all
$\delta\in (0,\sqrt k\,]$ and $R_{\rm R}(\sqrt k\,)=0.$ Moreover,
$R_{\rm R}(\delta)<R_{\text{LP}}(\delta)$ for most values of $\delta$
except for values in a small neighborhood of $\sqrt k.$ 
The intersection point $\delta_\ast$ of the curves $R_{\text R}$ and
$R_{\text{LP}}$ is given in the following table.

\smallskip\begin{tabular}{cccccc}
$k$&2&3&4&5&10\\
$\delta_{\ast}$&1.37&1.717&1.992&2.231&3.161
\end{tabular}

\smallskip The behavior of the bounds for $k=3$ is shown in the figure.

\subsection*{2. Proof.} The proof combines the isometric embedding 
of $G_{k,n}$ in $S_r$ with an application of Blichfeldt's density method
similar to the arguments of Rankin \cite{ran55}.
The intuition behind this method is as follows. 
Consider an $(M,\delta)$ code
$\cG\subset G_{k,n}$. Denote by $
B_\delta=B_\delta(x)$ a metric ball in $G_{k,n}$
with center at $x.$ Open 
balls of radius $\delta/2$ centered at code points do not intersect,
so no point of $G_{k,n}$ can be contained in more than one such ball. 
The idea is to extend the radius $\delta/2$ to some radius $\rho$ so that
while one point can belong to several balls, we can control the
way the balls intersect and use some type of the volume argument
to derive an upper bound on $M$. This idea, first suggested by Blichfeldt,
can be viewed as a precursor to the  
well-known Elias bound of coding theory (see e.g., \cite[p.61]{Lin99a}). 

Formally this idea is developed as follows.
Under the mapping $\Phi: G_{k,n}\to S_r$ an $(M,\delta)$ code $\cG$ is mapped
to a spherical code $\cA$ with minimum angular distance $2\alpha,$
where $\delta=\sqrt 2 r\sin\alpha.$ Let $\beta$ be the angle given by 
$\sin\beta=\sqrt 2\sin \alpha$ and let $\rho=\sqrt2 r\sin\nicefrac\beta 2.$ 
We compute
  \begin{equation}\label{eq:rho}
 \rho=r\sqrt{1-\sqrt{1-\sin\beta^2}}=r\sqrt{1-\sqrt{1-\frac{\delta^2}{r^2}}}.
  \end{equation}
Let $p\in \cG$ be fixed and let $q\in G_{k,n}$ be a point (plane) whose
principal angles to $p$ are given by $\theta=(\theta_1,\dots,\theta_k).$
Let $d=\|\sin\theta\|$ be the value of the distance between $p$ and $q.$
Consider the function on $G_{k,n}$ defined by
   $$
     \tau_p(q)=\begin{cases}
             \frac{2\cos\beta}{r^2\sin^2\beta}(\rho^2-d^2)
                   &\text{if }d\le\rho\\
             0&\text{if }d>\rho.
         \end{cases}
  $$
In other words, $\sigma_p(q)$ can be viewed as a ``density'' defined on
the metric ball $B_\rho\subset G_{k,n}$ with center at a point $p\in \cG$ and
radius $\rho.$ It depends only on the distance to the
center (is spherically symmetric).

Let us project the sphere $S_r$ radially on the unit sphere in 
$S\in \reals^{(n-1)(n+2)/2}$ and denote the image of the code $\cA$
by $\cC.$ Applying $\Phi$ followed by the projection to 
the ball $B_\rho$ with center at $p$ transforms it into
a cap on $S$ with angular radius $\beta$ and center at
$x=(1/r)\Phi(p)$ on the surface of the sphere.
The linear radius of the cap equals $P=2\sin\nicefrac\beta2.$
Letting $q$ be a plane at distance $d$ from $p$,
we observe that the distance between $x$ and $z=\Phi(q)$ equals $s=\sqrt 2d/r$.
The function $\tau$ induces a function $\sigma$ on this cap 
defined with respect to $x$ by
   $$
     \sigma_{x}(z)
      =\frac{\cos\beta}{\sin^2\beta}(P^2-s^2)
           $$
for $s\le P$ and $\sigma_x(z)=0$ otherwise.
A point $z$ can belong to several caps with centers at points
of the code $\cC$.
The following lemma, whose proof is included for completeness, is due to
\cite{ran55}.
\begin{lemma} 
For any point $z\in S$, its total density satisfies
  $$
    \sum_{x\in C} \sigma_x(z)\le 1.
  $$
\end{lemma}
\begin{proof}
Let $\cC\subset S$ be a code with distance $\tilde\delta$ and
let $z\in S$ be a point. Denote by $x_1,\dots x_m\in \cC$ the code
points whose distance to $z$ is at most $P$
and let $d_1,\dots, d_m$ be the values of these distances.
We have
  $$
      \frac12 m(m-1)\tilde\delta^2\le 
        \frac12
        \sum_{i=1}^n\sum_{j,k=1}^m(x_{ij}-x_{ik})^2=\sum_{i=1}^n\Big\{m\sum_j 
        x_{ji}^2-\Big(\sum_k x_{ki}
        \Big)^2\Big\}
        $$
          $$
     =\sum_{i=1}^n\Big\{m\sum_{j=1}^m x_{j,i}^2-\Big(\sum_{j=1}^m x_{ji}
        \Big)^2\Big\}
     $$
  $$
  =m\sum_j (1-x_{j1})^2-\Big(m-\sum_j x_{j1}\Big)^2+
                \sum_{i=2}^n\Big( m\sum_j x_{ji}^2-\Big(\sum_{j} x_{ji}\Big)^2
                 \Big)
  $$
Without loss of generality let $z=(1,0,0,\dots,0).$
Since
   $$
     d_j^2=(1-x_{j1})^2+x_{j2}^2+\dots+x_{jn}^2=2(1-x_{j1})
   $$
we obtain the inequality
    $$
     \frac12 m(m-1)\tilde\delta^2\le m\sum_j d_j^2-
              \Big(\sum_j \frac{d_j^2}2 \Big)^2-\sum_{i=2}^n\Big(\sum_j
          x_{ji}\Big)^2
     $$
which implies
  \begin{equation}\label{eq:lr}
     \Big(\sum_jd_j^2\Big)^2-4m\sum_{j} d_j^2+2m(m-1)\tilde\delta^2
        \le 0.
  \end{equation}
Let $\alpha_z=\sum_{j=1}^m \sigma_{x_j}(z).$ We have
  $$
    \alpha_z=\frac{\cos\beta}{\sin^2\beta}(mP^2-\sum d_j^2)
  $$
Then
   $$
     \sum d_j^2=4m \sin^2\nicefrac\beta2-\alpha_z\sin\beta\tan\beta=
       4\sin^2\nicefrac\beta2\Big(m-\frac{1+\cos\beta}
            {2\cos\beta}\alpha_z\Big).
   $$
Using this in (\ref{eq:lr}) we obtain
  $$
    16\sin^4\nicefrac\beta2(m-\nicefrac12(1+\sec\beta)\alpha_z)^2-
        16m\sin^2\nicefrac\beta2(m
              -\nicefrac12(1+\sec\beta)\alpha_z)
   $$
     $$
     +2m(m-1)\tilde\delta^2\le 0.
  $$
This inequality reduces to $4m(1-\alpha_z)\ge\alpha_z\tan^2\beta$ which
implies the claim of the lemma.
\end{proof}

Therefore also for any point $q\in G_{k,n}$
    $$
      \sum_{p\in C}\tau_p(q) \le 1.
     $$
Let $m(B_\rho)$ be the total mass of the ball computed with respect
to the density $\tau.$ From the last inequality we obtain
   \begin{equation}\label{eq:dens}
      M m(B_\rho)\le \Vol(G_{k,n})
  \end{equation}
where $\Vol(G_{k,n})$ is the total volume of the space.
Let $\mu(B_{\rho})= m(B_\rho)/\Vol(G_{k,n})$ be the normalized mass.
We assume that $k<n/2$.
The volume form on (the open part of) $G_{k,n}$ induces a 
distribution on the simplex of principal angles 
$\Theta=\{(\theta_1.\dots,\theta_k): \nicefrac\pi2>\theta_1>\dots>\theta_k>0\}$
given by
    $$
   \omega_{k,n}=K(k,n)\prod_{i=1}^k (\sin\theta_i)^{n-2k}
      \prod_{1\le i<j\le k}(\sin^2\theta_i-\sin^2\theta_j)
        d\theta_1\dots d\theta_k,
   $$
where $K(k,n)$ is a constant chosen from the
normalization condition $\int_{G_{k,n}} \omega_{k,n}=1$
(see, e.g., \cite{jam54}).
Then
  $$
    \mu(B_\rho)=\int_{\theta: \|\sin\theta\|\le \rho} 
      \tau(\|\sin\theta\|)\omega_{k,n}.
  $$
Asymptotic evaluation of an integral very similar to this
one was performed in \cite{bar02a}. We state the result in the
following lemma whose proof is analogous to \cite{bar02a}.
\begin{lemma} Let $k$ be fixed and $n\to\infty.$ Then
$$
  \mu(B_\rho)=\Big(\frac \rho{\sqrt{k}}\Big)^{nk+o(n)}.
$$
\end{lemma}
\nd Substituting the last formula in (\ref{eq:dens}) and taking logarithms
we obtain
  $$
    \frac1n\ln M\le -k \ln \frac\rho{\sqrt k}+o(1).
  $$
Finally, using (\ref{eq:rho}) and noting that
$r\to \sqrt k$ as $n\to\infty$, we obtain the bound of the theorem. \qed

\medskip
The result of Theorem 1 can be also extended to the complex Grassmannian
space similarly to an extension to this case of the bounds 
(\ref{eq:gv-hamm}) in \cite{bar02a}. 
These estimates can also be extended to the quaternionic case
according to the results of \cite{vre84}.

\providecommand{\bysame}{\leavevmode\hbox to3em{\hrulefill}\thinspace}

\end{document}